\documentstyle[prb,aps,psfig]{revtex}
\begin{document}
\title{Gap deformation and classical wave localization in disordered
two-dimensional  photonic band gap materials}
\author{
E.~Lidorikis$^1$, M.~M.~Sigalas$^1$, E.~N.~Economou$^2$ and 
C.~M.~Soukoulis$^{1,2}$
}
\address{
$^1$Ames Laboratory-U.S.~DOE and Department of Physics and Astronomy,   
Iowa State University, Ames, Iowa 50011 \\
}
\address{
$^2$Research Center of Crete-FORTH and Department of Physics,   
University of Crete,Heraklion, Crete 71110, Greece \\
}
\author{\parbox[t]{5.5in}{\small
By using two {\em ab initio} numerical methods we study the effects
that disorder has on the spectral gaps and on wave localization in 
two-dimensional photonic band gap materials. We find that there are
basically two different  responses depending on the lattice realization 
(solid dielectric cylinders in air or {\em vise versa}), the 
 wave polarization, and  the particular form under which disorder is 
introduced. 
Two different pictures for the photonic states are employed, 
 the ``nearly free'' photon and  the ``strongly localized'' photon.
These originate from the two different mechanisms responsible for the
formation of the spectral gaps, ie. multiple scattering and single
scatterer resonances, and they qualitatively explain our results.
\\ \\
PACS numbers: }}
\maketitle
\normalsize 

\section{Introduction}

Electromagnetic waves traveling in periodic dielectric structures
will undergo multiple scattering. For the proper structural parameters 
and wave frequencies, all waves may backscatter coherently;
the result is total inhibition of propagation inside the structure. 
Such structures are called photonic band gap (PBG) materials 
\cite{soukoulis,joannopoulos} or
photonic crystals, and the corresponding frequency ranges, for which 
 propagation is not allowed, photonic band gaps or stop bands.
PBG materials can be  artificially made in one, two, or three dimensions.
For example, a periodic lattice of dielectric spheres embedded in a 
different  dielectric  medium would work as a three-dimensional PBG
material, for the proper choice of lattice symmetry, dielectric contrast,
and sphere volume filling ratio. In two dimensions, a periodic array
of parallel, infinitely long, dielectric cylinders could work as a 
two-dimensional PBG material, prohibiting  propagation in a direction 
perpendicular to the cylinders' axis for some frequency range(s).
The absence of optical modes in a photonic band gap is often  considered
as analogous to the absence of electronic energy eigenstates in the 
semiconductor energy gap. The ability of PBG materials to modulate
electromagnetic wave propagation, in a similar way semiconductors  
modulate the electric current flow, can have a profound impact in many 
areas in pure and applied physics. It is then of fundamental importance
to study the effects of disorder \cite{fan,sigalas} 
on the transmission properties of  such materials.

Besides  the  non-resonant, macroscopic  Bragg-like multiple scattering, 
there is also a second, resonant mechanism, that contributes to the
formation of the spectral gaps. This is \cite{ho,kafesaki,lidorikis}
  the excitation
of single scatterer Mie resonances \cite{mie}. 
In a previous publication \cite{lidorikis} it was 
shown that for two-dimensional PBG materials, for the $E$ polarization 
scalar wave case (electric field parallel to the cylinders' axis),
these Mie resonances are analogous to the electronic orbitals in
semiconductors. The idea of the linear combination of atomic orbitals 
(LCAO) method was extended to the classical wave case as a linear
combination of Mie resonances (LCMR), leading to a successful tight-binding
(TB) parameterization for  photonic band gap materials.  
This moves the picture for the photon states, from a one analogous
to the nearly free electron model, to the one analogous to the strongly
localized electron whose transport is achieved only by hopping
(tunneling) from atom to atom. Depending then on which mechanism
is dominant for the formation of the photonic gaps, we expect
different changes to the system's properties when disorder is introduced.
If the  Bragg-like multiple scattering mechanism is the dominant one,
the photonic gaps should close quickly with increasing disorder, while
if it is the excitation of Mie resonances, the photonic gaps should
survive large amounts of disorder, in a similar way the electronic 
energy gap survives in amorphous silicon. 

In this paper we will use two {\em ab initio} numerical methods to study the
effects of disorder on photonic gap formation and wave localization
in two-dimensional PBG materials. 
The first is the finite difference time domain (FDTD) spectral method
\cite{yee,taflove},
from which we obtain the photonic density of states for an infinite,
disordered PBG  material, and the second is the transfer matrix technique
\cite{pendry}, 
from which we obtain the transmission coefficient for a wave incident
onto a finite slab of the  disordered PBG  material. From the transmission 
coefficient we can obtain the localization length for the photonic states
of the disordered material \cite{sigalas}. The study will be on both 
PBG material realizations (solid high dielectric cylinders 
in air and cylindrical  air holes
 in high dielectric), for both wave polarizations,  and it will 
incorporate three different disorder realizations: disorder in position,
radius, and dielectric constant (these systems, though, will still be 
periodic on the average). We  will find that only the case 
of solid dielectric cylinders in air with the wave $E_z$-polarized 
exhibits the 
behavior
expected from the strongly localized photon picture, while for
all other cases, the nearly free photon picture seems to be the dominant one.

\section{Numerical methods}

Electromagnetic wave propagation in lossless composite dielectric
media is described by Maxwell's equations
\begin{eqnarray}
\label{disord_eq1}
\mu \frac{\partial \vec{H}}{\partial t}  = 
 -\vec{\nabla} \times \vec{E}, & \hspace{1.5cm} &  
\epsilon(\vec{r})\frac{\partial \vec{E}}{\partial t}  = 
\vec{\nabla} \times \vec{H},
\end{eqnarray}
where the dielectric constant $\epsilon(\vec{r})$ is a function of position.
In two dimensions, the two independent wave polarizations are decoupled.
We assume the variation of the dielectric constant, as well as the
propagation direction, along the $xy$ plane, and so, the cylinders along
the $z$ axis. One of the polarizations 
is with the electric field parallel to the $z$ axis and the magnetic
field on the $xy$ plane ($E_z$ or TM polarized) 
and obeys a scalar wave equation. The other one with the magnetic field 
parallel to the $z$ axis and the electric field on the $xy$ plane 
($H_z$ or TE polarized) and obeys a vector wave equation. 

The first method we will use, to study disordered PBG materials, is
the FDTD spectral method \cite{chan,sakoda}.
In our FDTD scheme, we first discretize the $xy$ plane into a fine 
uniform grid. Each grid point is centered in a unit cell which is further
discretized into a 10$\times$10 subgrid, on which an arithmetic average 
of the dielectric constant is performed. In our problem we will assume
dispersionless and lossless materials. For the $E_z$ polarization case we
 define the electric field on this grid and the magnetic field 
on two additional grids, one  tilted by $(d/2,0)$,
 on which $H_y$ is defined,
and one tilted  by $(0,d/2)$, on which we define $H_x$.
 $d$ is the side of the 
grid cell.
The corresponding finite-difference equations for the space
derivatives that are used in the curl operators are then central-difference
in nature and second-order accurate. The electric and magnetic fields are
also  displaced  in time  by  a half time step $\Delta t/2$, 
resulting in a ``leapfrog'' arrangement and  central-difference equations 
for the time derivatives as well. If one initialize the electric and
magnetic fields at $t=t_0$ and $t=t_0+\Delta t/2$ respectively, then
updating the values of the electric field for each grid point $(i,j)$ at
$t=t_0+\Delta t$ is done by
\begin{eqnarray}
\label{disord_eq2}
E_z\vert_{i,j}^{t_0+\Delta t}  =  E_z\vert_{i,j}^{t_0} & + & \frac{\Delta t}
{d\ \epsilon_{i,j}} ( H_y\vert_{i+1/2,j}^{t_0+\Delta t/2} -
H_y\vert_{i-1/2,j}^{t_0+\Delta t/2}- \nonumber \\ 
 & - & H_x\vert_{i,j+1/2}^{t_0+\Delta t/2}+H_x\vert_{i,j-1/2}^
{t_0+\Delta t/2} )
\end{eqnarray}
where $\epsilon_{i,j}$ is the averaged dielectric constant for the grid
point $(i,j)$. Similar equations follow for  updating the magnetic field
components at $t=t_0+3\Delta t/2$, then again Eq.~(\ref{disord_eq2})
for $E_z$ at $t=t_0+2\Delta t$ etc. This way the time evolution of
the system can be recorded. For numerical stability and  good convergence 
the number of grid points per wavelength $\lambda/d$  must be at least 20,
 and also $\Delta t \leq d/\sqrt{2}c$, where $c$ the speed of light in vacuum. 
Similar equations, with the roles of the 
electric and magnetic fields interchanged, apply for the $H_z$ polarization
case.
 
In order to find the eigenmodes of a particular periodic (or disordered) 
 system, we first initialize the electric and magnetic fields in the
unit cell (or a suitable supercell) using periodic boundary  
conditions: $\vec{E}(\vec{r}+\vec{a})=e^{i\vec{k}\vec{a}}\vec{E}(\vec{r})$ 
and similarly
for  $\vec{H}(\vec{r})$, where $\vec{k}$ is the corresponding  
Bloch wave vector
and $\vec{a}$ the lattice vector. These fields must  have nonzero 
projections with the modes in search. We choose a superposition of
Bloch waves for the magnetic field and set zero the electric field:
\begin{eqnarray}
\label{disord_eq3}
\vec{H}(\vec{r})=\sum_{\vec{g}} \hat{v}_{\vec{g}}
e^{i(\vec{k}+\vec{g})\vec{r}+i\phi_{\vec{g}}}, & \hspace{1cm} &
\vec{E}(\vec{r})=0,
\end{eqnarray}
where $\phi_{\vec{g}}$ is just a random phase and the unit vector $\hat{v}$ 
is perpendicular to both $\vec{E}$ and $(\vec{k}+\vec{g})$, ensuring that
$\vec{H}$ is transverse and that $\vec{\nabla}\vec{H}=0$. 
Once the initial fields are 
defined, we can evolve them in time using the ``leapfrog'' difference
equations, while recording the field values as a time series for some
sampling points.  As the electric fields ``builds'' up, some particular
modes dominate while most are depressed, reflecting the underline
lattice symmetries. Here we record only the $E_z$ field for the $E_z$ 
polarization case, and the $H_z$ field for the  $H_z$ polarization. 
At the end of the simulation, the time series are Fourier 
transformed back into frequency space, and  the eigenmodes 
$\omega(\vec{k})$ of the system appear as sharp peaks. 
The length of the simulation determines the 
frequency resolution while the time difference between successive recordings
determines the maximum frequency considered. This method scales
linearly with size: a larger system will still need the same number of
time steps for the same  frequency resolution, thus sometimes referred
to also as an ``order-N'' method \cite{chan}. 

Here we will use this method to  obtain the system's density of states
(DOS). If one chooses a large supercell instead of the unit cell, then
for each $\vec{k}$ point inside it's first Brillouin zone,
the Fourier transformed time series will consist of a number of peaks.
Adding all contributions from all $\vec{k}$'s will result to a smooth
function for the DOS. This is in contrast to older methods that where
using random fields as initial boundary conditions \cite{sigalas}. 
Random initial fields 
will  ensure the condition for nonzero projections to all of the system's 
eigenmodes, but in order to get coupled with them during ``built'' up,
a large simulation time  is
required. Furthermore, the produced DOS  is not a smooth function of frequency,
still consisting of a large collection of peaks, and thus being useful only
as an indication for the existence of spectral gaps. 
In our method, the underline symmetries of the modes are already in the 
initial fields and so they couple easier with them. Also, the larger the 
supercell, the smaller is its first Brillouin zone, and so the smaller the 
frequencies we initialize through the various $\vec{k}$. This is why 
we can get smooth results even for very low frequencies.
In Figs.~1 and 2, we show the calculated density of states for the case
of solid dielectric cylinders in air and cylindrical air holes
in dielectric respectively, both for a square lattice arrangement, and
 for both polarizations. Along with
them we also plot the corresponding band structure as obtained with
the plane wave expansion method. Our study is going to be based on these two 
photonic structures.

The second method we will use is the transfer matrix technique in
order to obtain the transmission coefficient for a wave incident
along the $xy$ plane on a slab (or a slice) of the photonic material. 
The slice is assumed uniform along the $z$ axis, and periodic along the $x$ 
direction through
application of periodic boundary conditions, while in the $y$ direction
it has a finite width $L$.
In this method one first constructs the transmitted waves at one side
of the slice and then integrates numerically  the time-independent 
Maxwell's equations to the other side. There, the waves are projected
into  incident and reflected waves, and so a value for the transmission
coefficient $T$ can be obtained. Here, we are interested in the 
wave localization in disordered photonic band gap materials, and in 
particularly in the localization length $\ell \sim -2L/\ln T$.

A few remarks about the results of this
method are in order. Waves with different
incidence directions will have different reflection and transmission
coefficients, so if one is looking for an average transmission, 
all directions should be included.  It is shown, however, that there
is also a large dependence on the surface plane  along which the
structure is cut. More specifically, a wave normally incident on 
a (1,0) surface will have different transmission characteristics
than a wave incident with 45$^{o}$ on a (1,1) surface. This is
because certain modes can not always get coupled with the incident wave.
One should then also average for the two different surface cuts,
otherwise it will not be a true average. This is shown in Figs.~3 and 4 
where we plot the (1,0) and the (1,1) cuts, each with both incidence 
directions (normal and 45$^{o}$ with respect to the surface) averaged.
We see that taken individually, none of them corresponds to the
true gaps as shown in Figs.~1 and 2, but rather, to wider and generally
displaced gaps. For example, in the $E_z$ polarization case in  the 
first spectral gap (Figs. 3a and 4a), with the (1,0) cut, the incident waves 
fail to couple with the  the {\bf M} modes of the first band, while 
with the (1,1) cut, the incident waves fail to couple with 
the {\bf X}  modes of the second band. 

This is expected to be lifted once disorder is introduced into our
system, since the sense of direction will be somehow lost. 
 Disorder can be introduced as a random displacement, a random
change in the radius, or, a random change in the dielectric constant
of the cylinders. It is not clear however what amount of disorder
would be needed for this. We repeated the calculations for small enough 
amounts of disorder so that the spectral gaps, as found from the
FDTD method, remain almost unchanged, for all three different 
disorder mechanisms. As seen in Figs.~3 and 4, indeed, in some cases  the
coupling is achieved. For example, for the first gap in the 
$E_z$-polarization case, with the (1,0) cut, the {\bf M} modes of the 
first band are now coupled
with the incident waves and appear in the transmission diagram. These
could be easily mistaken for disorder-induced localized states
entering the gap, but they are not, since for the values of disorder used, 
the first gap is virtually unchanged. On the other hand, with the (1,1)
cut, the coupling to {\bf X} modes of the second band is not yet achieved, 
still yielding a wrong picture
for the gap. Increasing the disorder further will eventually destroy
any sense of direction and there will be no distinction between the
two cases. Figs.~3 and 4 will be useful as a guide of which results can
be trusted and which can not, if one uses only one surface cut and
small values of disorder. As a general rule, we can deduce that the 
(1,0) cut should be used for the $E_z$ polarization case, while the
(1,1) cut would be better for the $H_z$ polarization case.

\section{Results and discussion}

We first looked into the spectral gaps' dependence  on disorder
using the FDTD spectral method. Our system consisted of a 8$\times$8
supercell, each cell discretized into a 32$\times$32 grid. We studied 
two systems:  a square lattice array of solid cylinders, with dielectric 
constant $\epsilon_a$=10, in air ($\epsilon_b$=1) with a filling
ratio $f$=0.28\%, and a square lattice array of air cylinders ($\epsilon_a$=1)
in  dielectric material $\epsilon_b$=10, with air filling ratio
$f$=0.71\%, as described in Figs.~1 and 2. We divided the 
supercell's first Brillouin zone into 10$\times$10 grid, which for
the irreducible part yields 66 different $\vec{k}$ points. 
For each particular disorder realization (i.e. disorder type) 
and disorder strength, 
we run the simulation for all these 66 $\vec{k}$'s. At each $\vec{k}$
however we use a different  disordered configuration 
(i.e. a different seed in the random number generator), and so a 
large statistical sample is automatically included in our result.
In each case the effective disorder is measured by the rms error
of the dielectric constant $<\epsilon>$, which is defined as 
\cite{chan,sigalas}
\begin{eqnarray}
\label{disord_eq4}
\epsilon^2=\frac{1}{N}\sum_{i=1}^{N}(\epsilon_i^d-\epsilon_i^p)^2,
\end{eqnarray}
where the sum goes over all $N=8\times8\times32\times32=65536$ grid points,
$\epsilon_i^d$ and $\epsilon_i^p$ are the dielectric constants at site
$i$ in the disordered and periodic case respectively, and $< \dots >$
means the average over different configurations (different $\vec{k}$'s
in our case). In both settings (dielectric cylinders in air and 
{\em vise versa}) the filling ratio of the high dielectric material 
is similar, and so $<\epsilon>$ is expected to have the 
same meaning and weight.

Four different disorder realizations are studied: 1) disorder in
position, without though allowing any cylinders to overlap
with each other, 2) disorder in position allowing 
cylinder overlapping to occur, 3) disorder in radius, and 4) 
disorder in dielectric constant (the last one only in the solid
cylinder case). For each different realization we consider various 
disorder strengths, and thus different effective disorders 
$<\epsilon>$,
for which we record the upper 
and lower gap edges for the first two photonic band gaps (if they exist).
Results are summarized in Figs.~5 and 6, for the solid and air
cylinder cases respectively. We note that 
 the $E_z$ polarization case for the solid cylinders is very different from
all other cases: the gaps  survive very large amounts of positional
disorder, especially if no overlaps are allowed. In fact, once the
disorder becomes large enough for overlaps to be possible,
the gap quickly closes, as shown in Fig.~5. The actual DOS graphs 
for the two different realizations of the positional disorder 
are shown in Fig.~7, for three different values of the effective
disorder. On the other hand, if the disorder is of the third or fourth 
kind, the gaps 
close very quickly, even for modest values of the effective disorder.

The picture is very different in the other cases, as seen in Fig.~6.
The effect of the positional disorder is the same, independent
of whether  overlaps are allowed or not. This is most clearly seen in Fig.~8, 
where the actual DOS graphs are plotted for the air
cylinder case  for both polarizations, and for both positional disorder
realizations. Allowing the  air cylinders to overlap,
though, means that  the connectivity of the background material will break. 
Our results, thus, indicate that there is no connection between the 
connectivity of the background material and the formation of the spectral
gaps in this 2D case.
 Most importantly, however, we note that the disorder in radius 
has a similar effect with that of the positional disorder in closing the gaps. 
In fact, it is also similar to the effect of the disorder in radius for 
the $E_z$-solid-cylinder case. 
So, in the case of air cylinders, the type of the disorder that is 
introduced into the system does not play a significant role, but rather,
it is only the effective disorder (measured through the dielectric constant's
error function) that determines the effect on the spectral gaps.
On the other hand, for the  $E_z$-solid-cylinder case, the type of disorder
plays a profound role: if the ``shape'' of the individual scatterer is
preserved, the gaps can sustain large amounts of disorder, while if it is not
preserved, the gaps collapse in a  manner similar to the air cylinder 
case.

We next go over  the localization length results, which were  obtained with
the transfer matrix technique. Here, our system consisted of a 3$\times$7
supercell (3 along the $x$ axis), with each cell discretized into a 
18$\times$18 grid 
(a small supercell
was used in order to ease the computation burden). In the $x$
direction we applied periodic boundary conditions, while in the $y$
direction the supercell was repeated 4 times, to provide a total length $L$
for the slab of $L$=28 unit cells. The structures studied are exactly the
same as described before. The lattice was cut along one only symmetry
direction, the (1,0), since for large disorders we expect all ``hidden''
modes to be coupled with the incident wave (in any case, we know from 
Figs.~3 and 4 which results can be completely trusted and which can
not). For each disorder realization and strength, we used 11 different
$\vec{k}$ values uniformly distributed between normal and 45$^o$ angle
incidence, and for each  $\vec{k}$ we used a different disordered 
configuration, so these will constitute our statistical sample.
For each $\vec{k}$ we find the minimum transmission coefficient
inside each gap, from which we find  the minimum localization length,
 and then average over all $\vec{k}$'s, 
ie. $\ell\sim-2L/<\ln T>$ (in the periodic case we first averaged over $T$
in order to correctly account for different propagation directions,
but in the highly disordered case it is not so much important any more, 
and so we  just average over the localization lengths).

Our results are shown in Figs.~9 and 10 (because of the small statistical
sample and the small supercell used, the data points appear very
``noisy'', especially for large disorders). We note here, as well, the distinct
difference between the $E_z$-solid-cylinder case for  positional
disorder and all other cases. Especially  for the first 
spectral gap, the localization length not only remains
unaffected by the disorder, but it even decreases (this is not
an artifact of the averaging procedure). The first
conclusion from this, is that the mechanisms responsible
for the gap formation in this case are unaffected by the presence 
of positional  disorder, and so they are definitely not macroscopic 
(long-range) in nature.
The fact that the localization length decreases, is attributed to the 
coupling of more  [1,1] symmetry modes with the incident wave 
as the disorder increases (they provide a smaller $\ell$ to
the average, as seen in Fig.~3a). This decrease should not be mistaken
for additional localization induced by the disorder (the
classical analog of Anderson localization in electrons), since the latter 
is macroscopic in nature, and does not apply for strongly 
localized waves.
The decrease in the localization length continues until a fairly
large disorder value, and then it increases to a  saturation value
(the dielectric error function can reach only up to some value  for
positional disorder).
This saturation value is higher for the case where overlaps are allowed,
but still is very small compared to other cases, so waves remain
strongly localized.

All other cases, on the other hand, show a common pattern of 
behavior: photon states become quickly de-localized with increasing
disorder. The localization length is increased, until the point where
the localization induced by the disorder becomes dominant. After 
this it starts decreasing, until finally it reaches some saturation point.
Note also that there is  an almost quantitative agreement between  
some cases that was not really  expected, eg. 
for the disorder in radius in  the first gap with the wave $E_z$-polarized,
for both lattice settings, as seen in Figs.~9a and 10a. Only the 
case of disorder in the dielectric constant seems to deviate,
having very quickly a very large effect, with the localization length 
 directly saturating to some constant value. So, for  air
cylinders in dielectric with any type of disorder, and for the 
$E_z$-solid-cylinder case with disorder that does not preserve the
scatterer's ``shape'', the behavior under disorder is similar.

All these results can be understood if we adopt two different ``pictures'' for 
the photon states, depending on which is the dominant 
 mechanism that is responsible  
for the formation of the spectral gaps in each case. The first is 
 the ``nearly free'' photon
picture, in which the gap forming mechanism is the non-resonant 
macroscopic Bragg-like
multiple scattering, while the second is the 
``strongly localized'' photon picture, in which the gap forming 
mechanism is the microscopic (short-range) excitation of single scatterer Mie 
resonances.

Sharp Mie resonances appear only for the solid cylinder 
case, and they can be thought as analogous to
the atomic orbitals in semiconductors.
Using this analogy, a tight-binding model, based on a linear 
combination of Mie resonances, was recently developed for the photonic
states in the $E_z$-solid-cylinder case \cite{lidorikis}. 
But if a tight-binding model
can give a satisfactory description of the photonic states, then it is
expected that certain behavioral patterns found in semiconductors 
should apply in our case too. So, positional disorder should have
a small effect on the gaps, in a similar way the energy gap survives
in amorphous silicon. Also, changing the scatterer should have a 
similar effect as changing the atoms in the semiconductor, yielding 
a large amount of impurity modes that quickly destroys  the gap. 
This pattern is definitely confirmed here for the $E_z$-solid-cylinder case.
In this case, multiple scattering and interference can 
only help to make the gaps wider, but are definitely not decisive on 
the existence of a gap. 

For the macroscopic Bragg-like multiple scattering mechanism, 
the lattice periodicity  is a very important factor for the existence 
of a spectral gap.  If it is destroyed, then coherence in the backscattered
waves will be destroyed, and so will the spectral gaps. It is of small 
consequence the exact way that the periodicity is destroyed, and so 
different disorder realizations will have similar effects. Also, since
the gaps close more easily, it will be easier to observe the localization
induced on the waves by the disorder itself, ie. the  classical 
analog of Anderson localization in electrons. All these are recognized
in the case of air cylinders in dielectric. 

Finally, in the $H_z$-solid-cylinder case, there were no gaps to
begin with, and so we can have no results about it. However, sharp
Mie resonances appear for this case as well, and if their excitation
was the dominant scattering mechanism, a gap would be expected here as well.
The difference with the $E_z$ is that the former is described by a
vector wave equation, while the latter by a scalar one (and thus
closer to the electronic case). The form of the wave equation must, 
then, be an important factor
in determining the relative strength of the two gap forming mechanisms.

\section{Conclusions}
We have shown that several results in periodic and random photonic
band gap materials can be understood in terms of two distinct photonic
states: (a) The ``local'' states, based on a single scatterer Mie
resonance, with the multiple scattering playing a minor role; these states
are more conveniently described in terms of an LCAO-type of approach and
are the analog of the $d$-states in transition metals. ``Local'' 
photonic states appear in the case of high dielectric cylinders 
surrounded by a low-dielectric host and for $E$-polarized waves.
(b) The ``nearly free'' photonic states, where Bragg-like multiple 
scattering is the dominant mechanism responsible for their appearance;
these states are more conveniently described in terms of a 
pseudopotential-type of approach and are the analog of $s$ (or $p$)
states in simple metals.

Each type of photonic states responds differently to the presence of 
disorder: For the ``local'' states case, the gap is robust as the 
periodicity is destroyed, and it is hardly affected by the disorder
as long as the identity of each individual scatterer is preserved;
however, if the shape, or other characteristics influencing the scattering
cross section of each individual scatterer, is altered by disorder, the 
gap tends to disappear. On the other hand, for the ``nearly free''
states case, the gap is very sensitive and tends to disappear easily
as the periodicity is destroyed.

\section{Acknowledgments} 
Ames Laboratory is operated for the U. S. Department of Energy by Iowa 
State University under contract No. W-7405-ENG-82. This work was supported
by the Director of Energy Research office of Basic Energy Science and 
Advanced Energy Projects. It was also supported by the Army Research Office, 
a E.U. grant, and a NATO
grant.


\begin{figure}
\psfig{figure=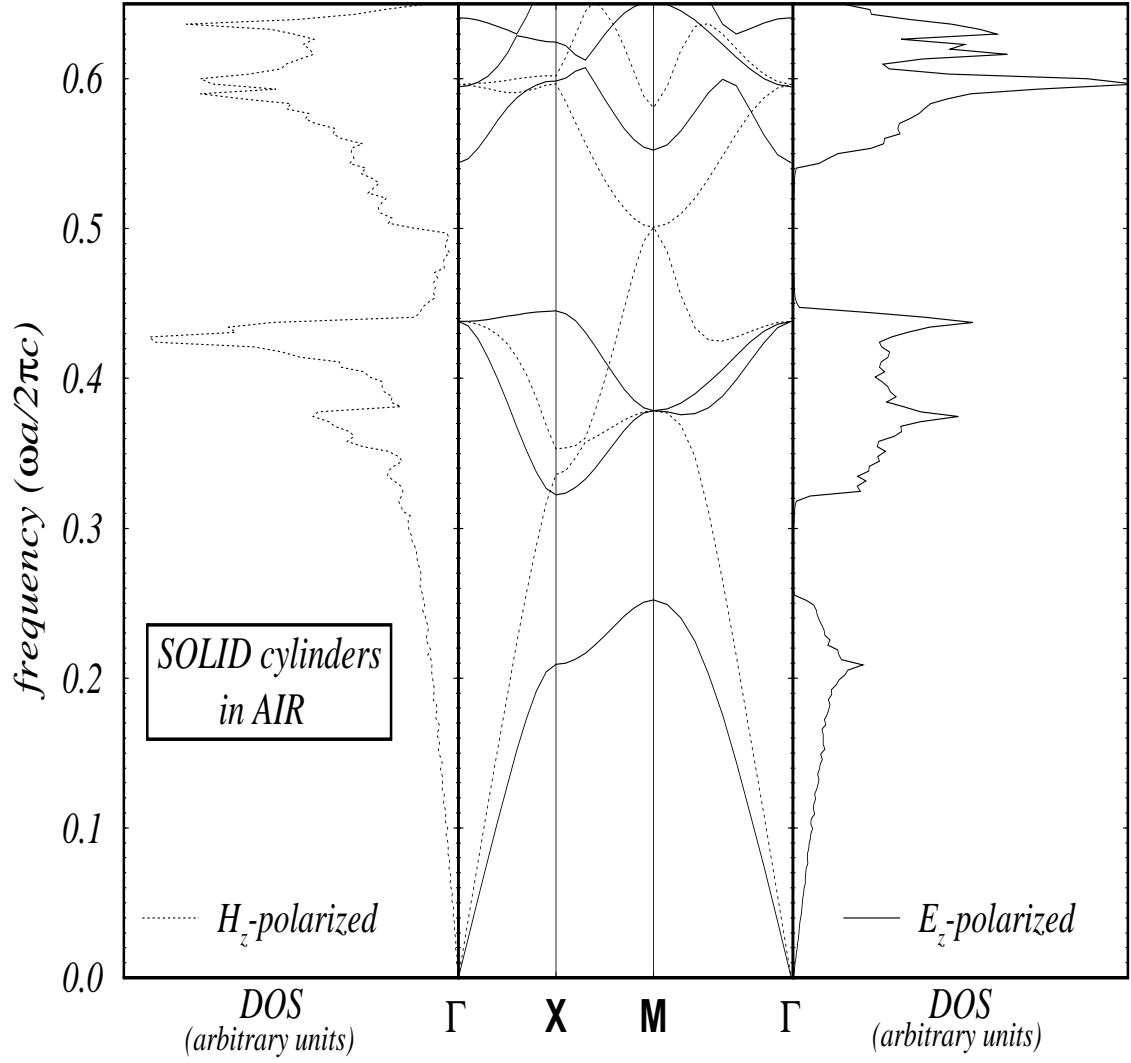,width=16cm,height=16cm,angle=270}
\caption{Band structure (obtained with a plane wave method) and density
of states (obtained with the FDTD spectral method) for a two-dimensional
square lattice array of dielectric cylinders $\epsilon_a$=10 in
air $\epsilon_b$=1, with a filling ratio $f\simeq 28$\%.}
\end{figure}

\begin{figure}
\psfig{figure=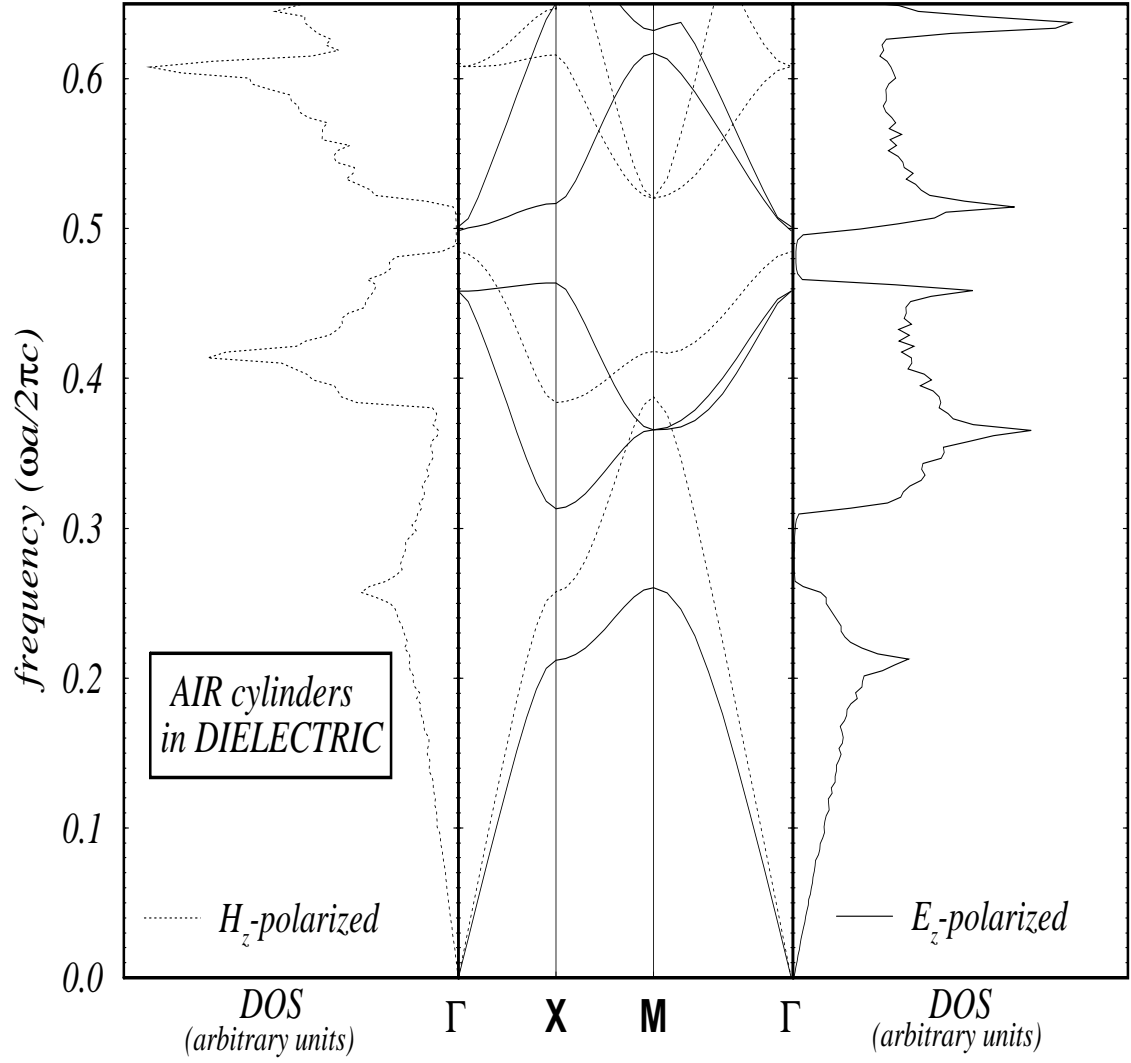,width=16cm,height=16cm,angle=270}
\caption{Band structure (obtained with a plane wave method) and density
of states (obtained with the FDTD spectral method) for a two-dimensional
square lattice array of air cylinders $\epsilon_a$=1 in
dielectric $\epsilon_b$=10, with air filling ratio $f\simeq 71$\%.}
\end{figure}

\begin{figure}
\psfig{figure=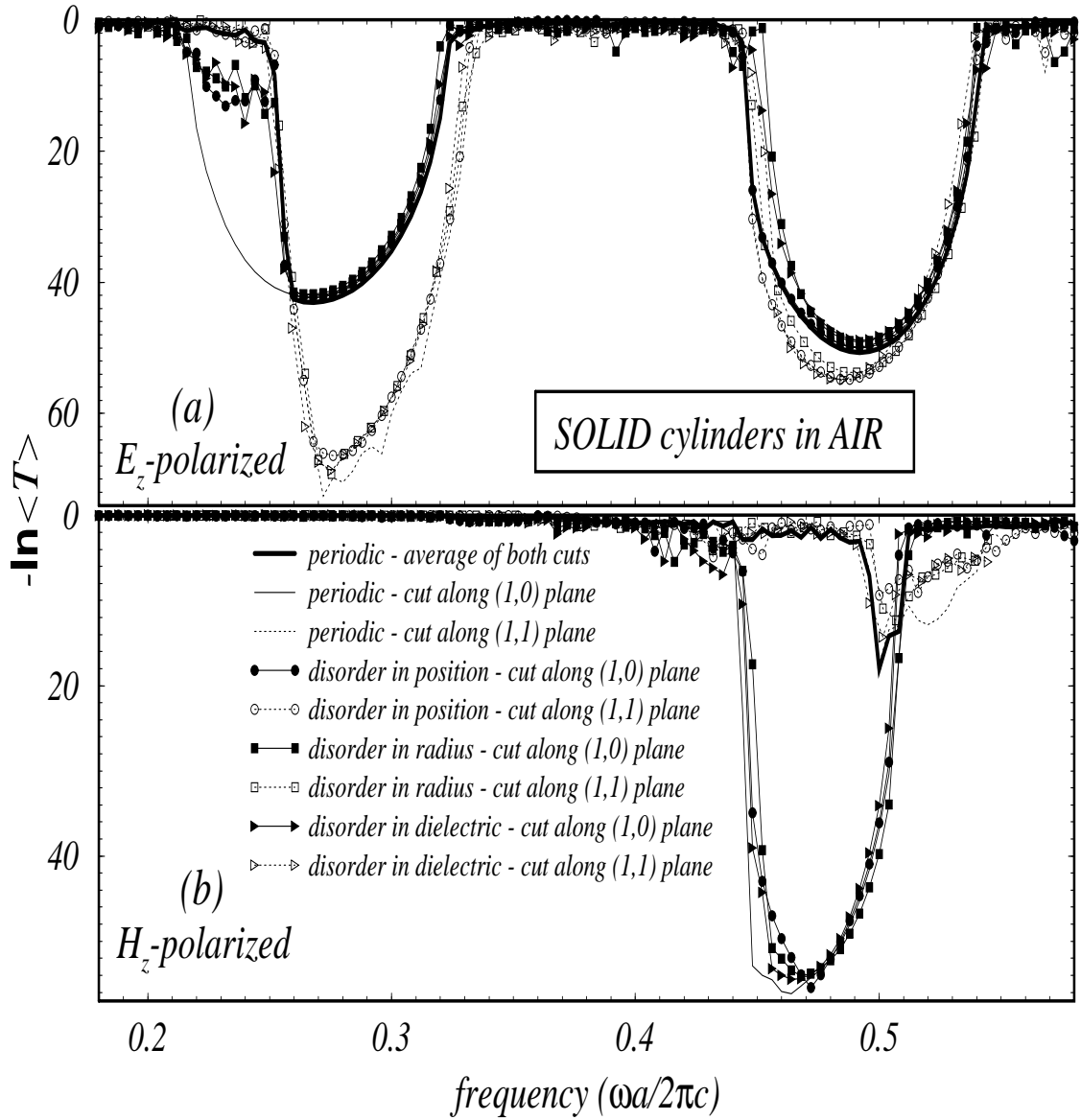,width=16cm,height=16cm,angle=270}
\caption{Transmission coefficient for the periodic, and weakly disorder,
system described in Fig.~1 (obtained with the transfer matrix technique).
Calculations are for two different surfaces along which the 
sample is cut. Effective disorders used (look Eq.~(\ref{disord_eq4})):
in position$\sim$1.3, in radius$\sim$0.5, and in dielectric$\sim$0.3.  
}
\end{figure}

\begin{figure}
\psfig{figure=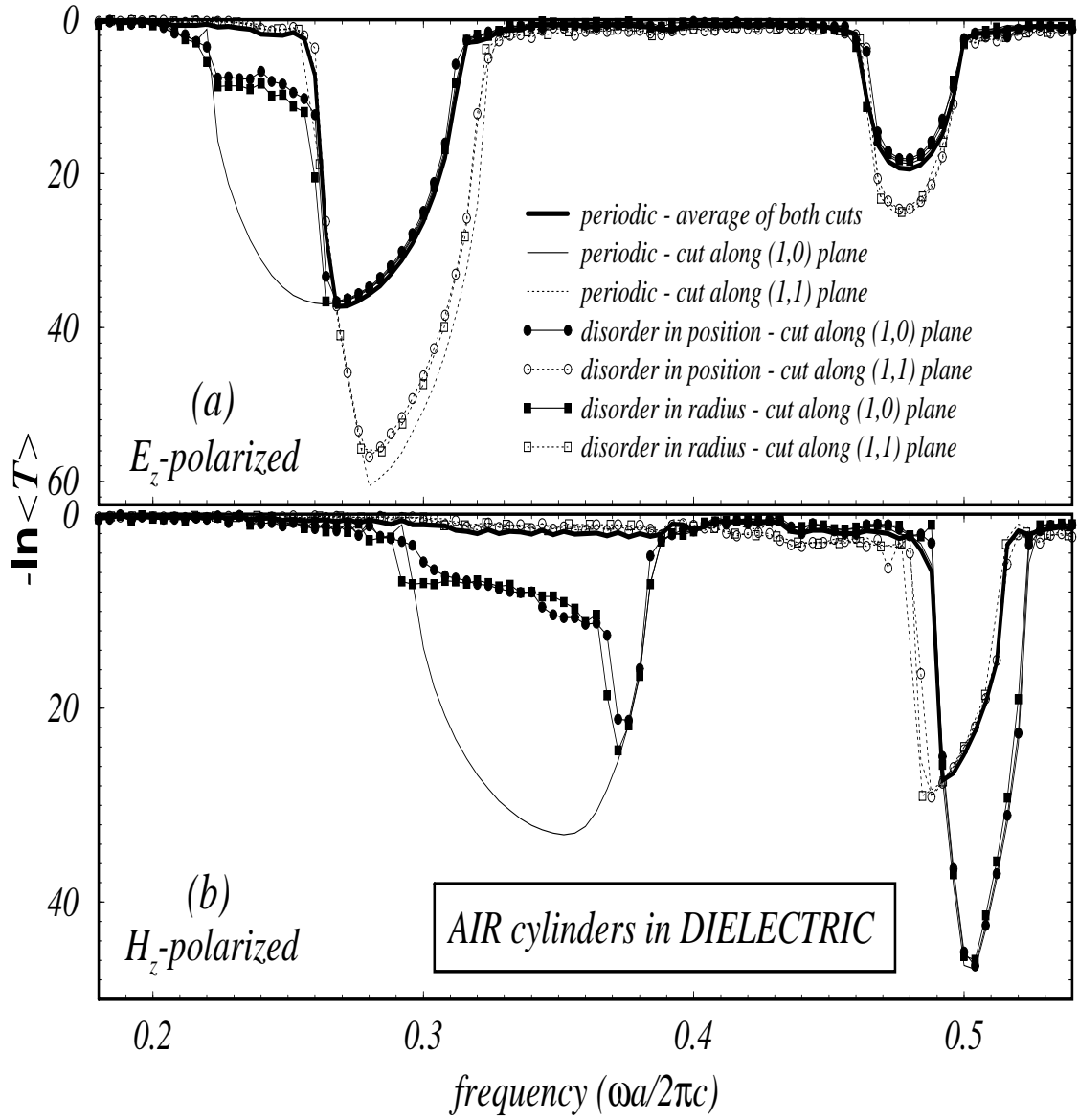,width=16cm,height=16cm,angle=270}
\caption{Transmission coefficient for the periodic, and weakly disorder,
system described in Fig.~2 (obtained with the transfer matrix technique).
Calculations are for two different surfaces along which the 
sample is cut. Effective disorders used:
in position$\sim$0.35, in radius$\sim$0.25}
\end{figure}

\begin{figure}
\psfig{figure=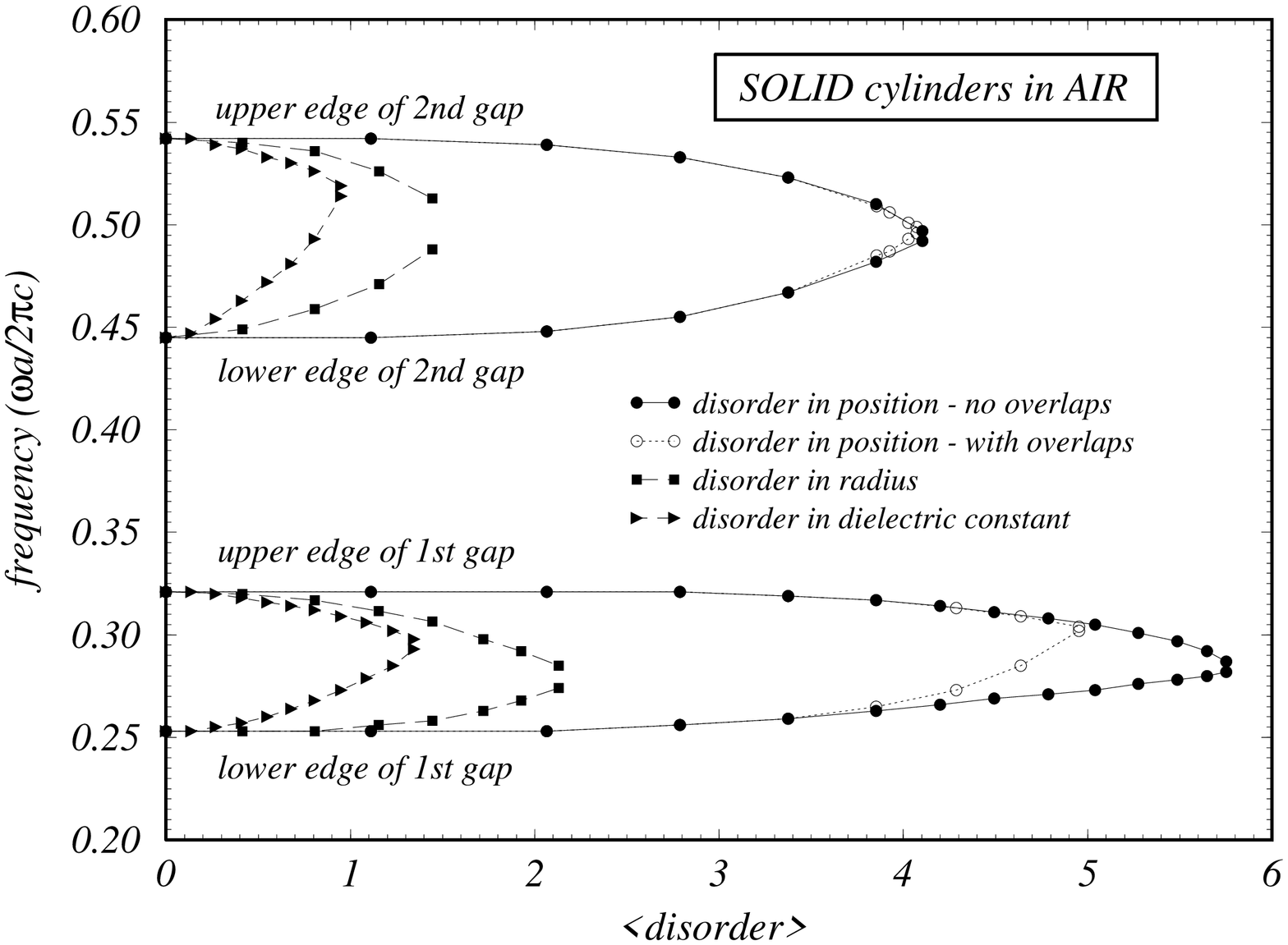,width=16cm,height=16cm,angle=270}
\caption{The edges of the photonic band gaps as a function of the
effective disorder $<disorder> \equiv <\epsilon>$
(as was defined in Eq.~(\ref{disord_eq4})), 
for the system described in Fig.~1. Four different disorder  
realizations are studied.}
\end{figure}

\begin{figure}
\psfig{figure=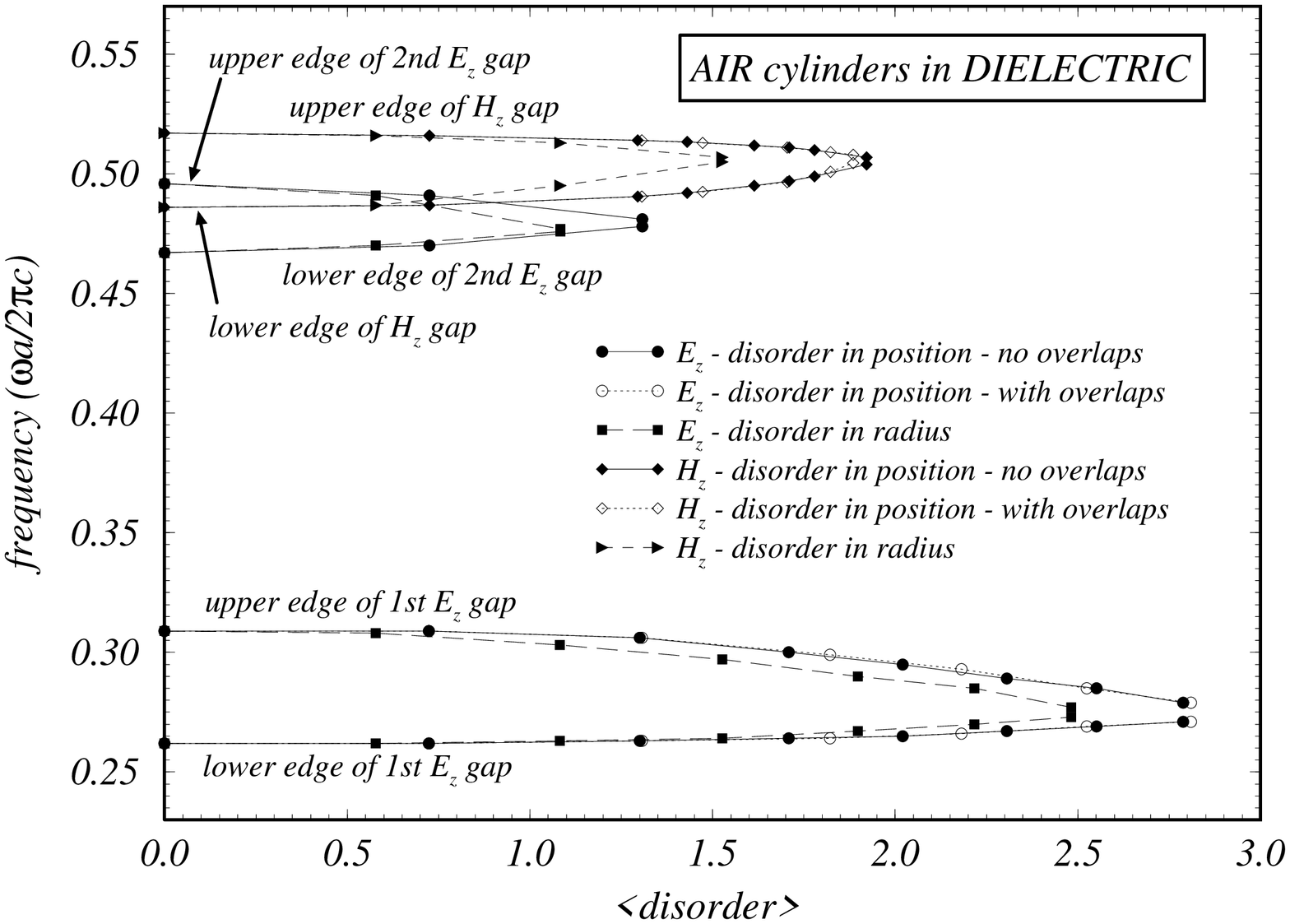,width=16cm,height=16cm,angle=270}
\caption{The edges of the photonic band gaps as a function of the
effective disorder $<disorder> \equiv < \epsilon>$
(as was defined in Eq.~(\ref{disord_eq4})), 
for the system described in Fig.~2.  Three different disorder  
realizations are studied.}
\end{figure}

\begin{figure}
\psfig{figure=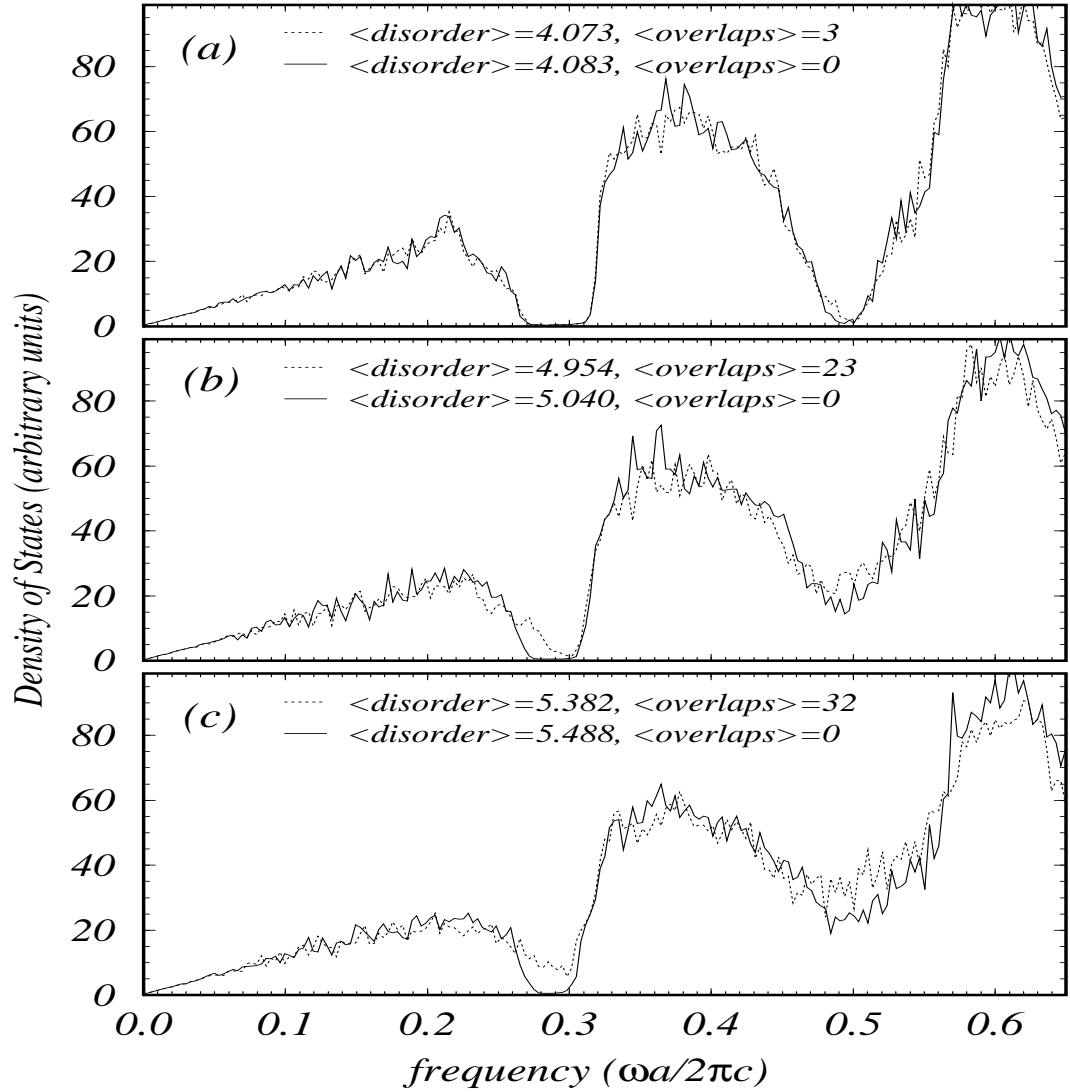,width=16cm,height=16cm,angle=0}
\caption{The density of states for the system of Fig.~1 with the $E_z$
polarization, for three different positional disorder strengths. The solid
line is when no scatterer overlaps are allowed, while the dotted line is
when scatterer overlaps are allowed. $<overlaps>$ is the average number 
of overlapping cylinders.}
\end{figure}

\begin{figure}
\psfig{figure=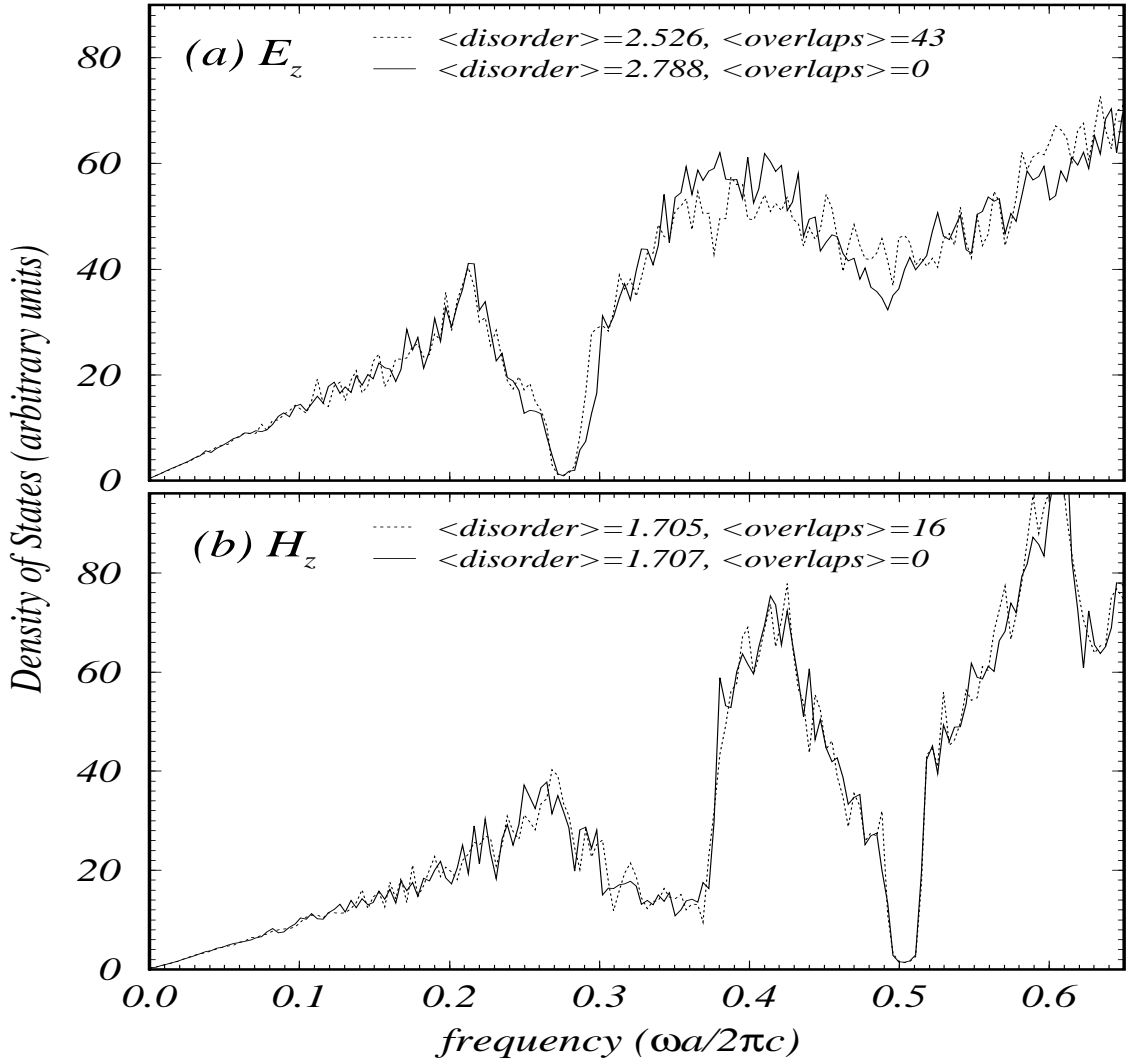,width=16cm,height=16cm,angle=0}
\caption{The density of states for the system of Fig.~2 for both field 
polarizations, for two different positional disorder strengths. The solid
line is when no scatterer overlaps are allowed, while the dotted line is
when scatterer overlaps are allowed.}
\end{figure}

\begin{figure}
\psfig{figure=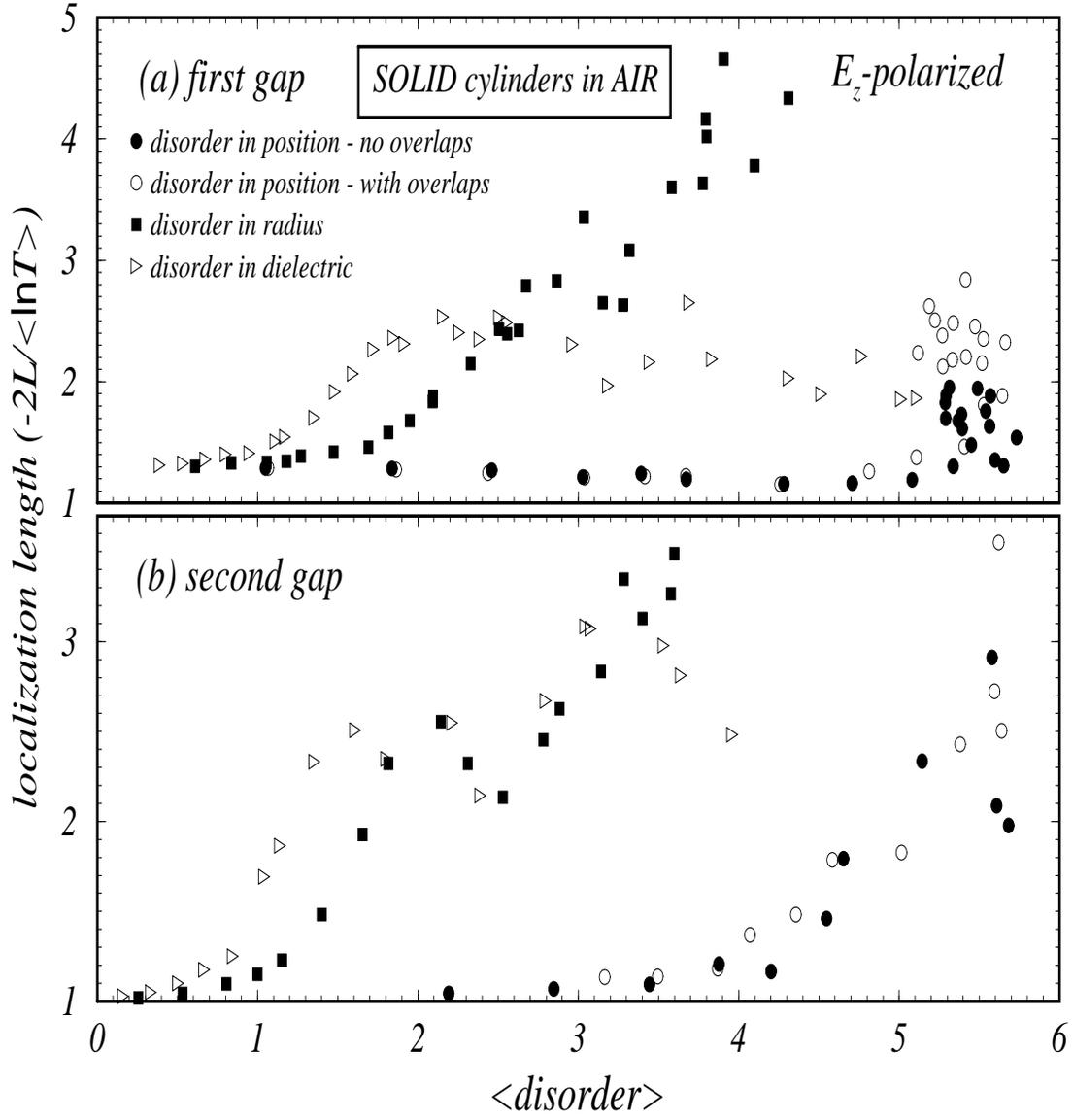,width=16cm,height=16cm,angle=270}
\caption{The localization length as a function of the effective disorder
for the system described in Fig.~1 with the $E_z$ polarization, for four
different disorder realizations.}
\end{figure}

\begin{figure}
\psfig{figure=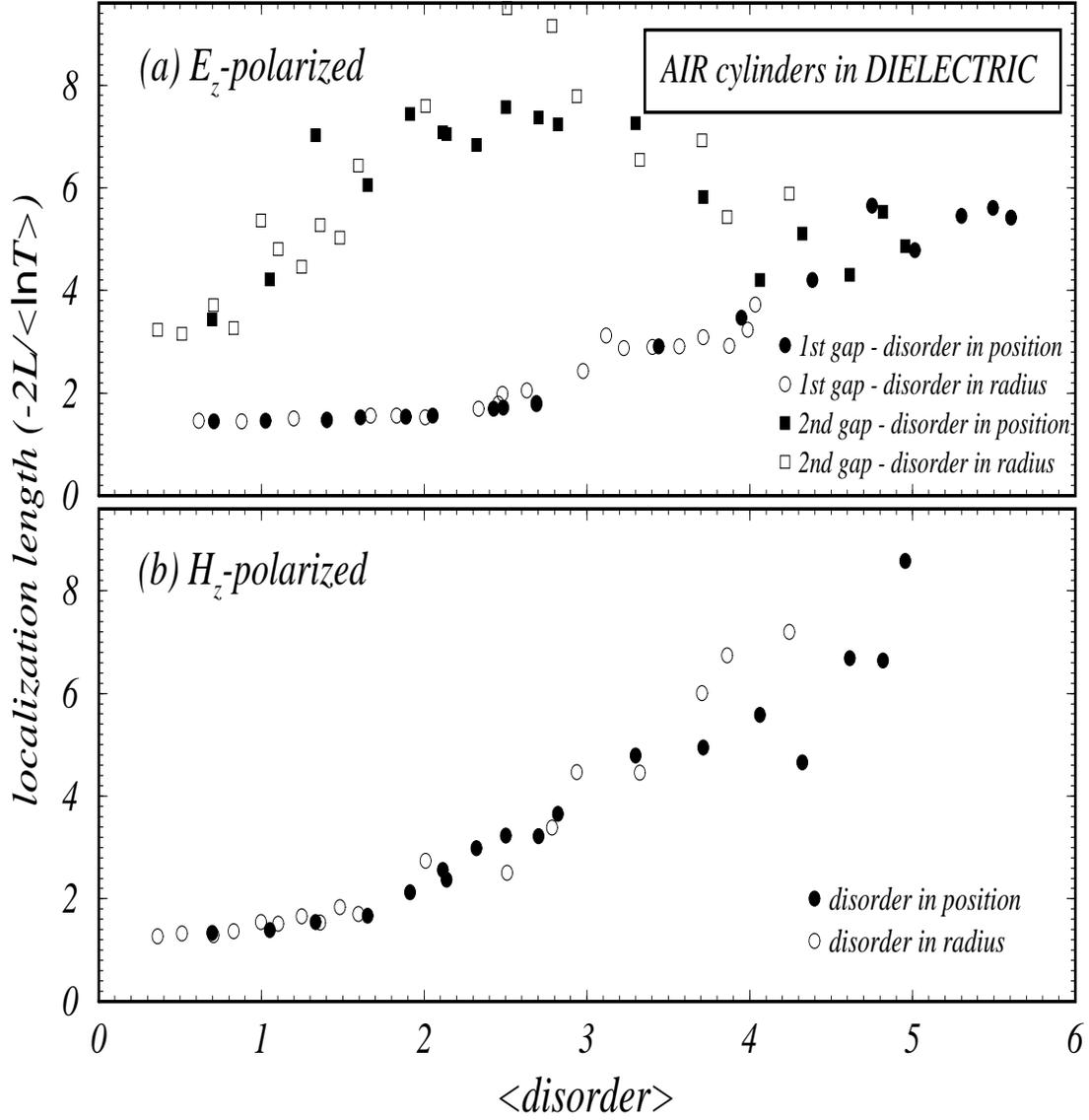,width=16cm,height=16cm,angle=270}
\caption{The localization length as a function of the effective disorder
for the system described in Fig.~2 for both field polarizations, for two
different disorder realizations.}
\end{figure}

\end{document}